\documentstyle[12pt,aaspp4]{article}
\def\D10{D_{10}}
\def\R9{R_{9}}
\def\t1{\tau_1}
\def\r6{\rho_6}
\def\rc6{\rho_{c,6}}
\def\Q52{Q_{52}}
\def\T9{T_9}
\def\Tc9{T_{c,9}}
\def\me{\mu_e}
\received{1995 March 7}

\begin{document}

\title{Induction of supernova-like explosions by $\gamma$-ray bursts
in close binary systems}

\author{Mordehai Milgrom and Vladimir V. Usov}

\affil{Department of Condensed-Matter Physics,
Weizmann Institute of Science, Rehovot
76100, Israel}

\authoremail{fnmilgrm@wicc.weizmann.ac.il, 
fnusov@weizmann.weizmann.ac.il}

\date{}

\begin{abstract}
We propose that a $\gamma$-ray burst in one member of a binary may induce
a supernova-like explosion of a close, white-dwarf companion.
 Such explosion might be brought about in rather light companions,
 which cannot undergo the standard accretion-induced explosion.
 This would give some GRB-associated supernova an appearance rather
unlike that of the typical Type I. 
GRB 980425, if indeed associated with 
SN 1998bw is too weak to have produced it through our proposed mechanism. 


\end{abstract}


\section{Introduction}
The prompt localization of gamma-ray bursts (GRBs) by
 BeppoSAX led to the discovery of long lived
GRB afterglows spanning the energy range from X-ray to radio
(\cite{Costa97}; \cite{vanParadijs97}; \cite{Frail97}), and of
associated host galaxies (\cite{Kulkarni98a}).  Detections of
absorption and emission features at high redshifts ($0.69\leq z \leq 
3.42$) in optical afterglows of GRBs and in their host galaxies (e.g.,
\cite{Kulkarni99}) have clearly tipped the scale in favour of the 
cosmological origin of GRBs sources
(\cite{Usov75}; \cite{vdBerg83}; \cite{Paczynski86}; \cite{Goodman86};
\cite{Eichler89}).

Despite such great advances
the exact nature of the GRB progenitors is still unknown. Several
currently popular models posit as the energy-releasing event:
coalescence of two neutron stars (\cite{Blinnikov84}; \cite{Paczynski86};
\cite{Eichler89}), the collapse of a massive star (\cite{Woosley93};
\cite{Paczynski98}); or the formation of a millisecond pulsar with
extremely strong magnetic field $(\sim 10^{15}-10^{16}$ G)
(\cite{Usov92}; \cite{Thompson93}; \cite{Blackman96};
\cite{Katz97}; \cite{Kluzniak98}; \cite{Vietri98}).
\par
The energetic ejecta of GRBs may affect their close surrounding in various
observable ways. For example, the reprocessing of some of the GRB
energy in the atmosphere of a companion might produce an optical 
afterglow, as discussed, e.g., by \cite{London83} and by  \cite{Melia86}
for galactical GRBs,  and recently by  \cite{bp98} for GRBs at
cosmological distances. In this Letter, we discuss another 
possible interaction of the GRB ejecta with a companion:
Some GRB explosions might occur in a close binary companionship 
with a white dwarf (WD). This, in fact, is a prerequisite in the GRB model of
Usov (1992). The interaction of the ejecta with the 
companion then may induce its explosion and appearance of a 
supernova-like phenomenon. The recently claimed supernovae in association 
with GRBs (e.g., \cite{Wheeler99}) are unlikely to
have been produced in this way. In \S 2 we consider the process of
induced explosion, and estimate the parameter values needed to actuate it.
In \S 3 we discuss possible observational consequences and other 
pertinent issues.


\section{Induction of white-dwarf explosions
by GRBs in close binaries}
\par
The observed fluxes of GRBs at cosmological distances
 imply total radiation energy output per unit solid angle within the
beam of radiation of $Q_{\gamma} \sim 10^{51} -10^{52}$ ergs/st
 on a time scale of seconds. The total angular energy
output, $Q$, is even larger,
as only part of it is converted into the observed radiation.
If the GRB source has a close binary companion that lies within
the main beam, a
powerful flux of energy impacts the latter.
 On near enough a companion the effects may be staggering. 
We consider such possible effects, and, in particular, the
possibility that the impact can induce a supernova-like 
explosion of a white-dwarf companion. The energy 
that falls on a unit area of the surface of the secondary that
is within the GRB beam is $q\simeq Q /D^2$, where
 $D=10^{10}\D10$~cm is the binary distance. 
The chances of an induced explosion are maximized when the 
companion is fully within the GRB beam, as we assume in
our discussion below. In this case the total energy that hits the 
companion is $\Delta Q\simeq \pi R_s^2q$, 
where $R_s=10^9R_9$~cm is the radius of the companion.
\par
Our proposal applies more generally, but, for the sake of 
concreteness, we consider system parameters that are natural
in the GRB model that
involves a strongly magnetized millisecond pulsar produced by
accretion-induced collapse of a white dwarf in a close binary
(\cite{Usov92}).  An inherent feature
of this model is that GRBs occur in binaries.
The GRB progenitor is a strongly magnetized white dwarf
with a mass near the Chandrasekhar limit,
$\sim 1.4M_\odot$. The secondary is a white dwarf with
a mass $M_s\simeq 0.3-0.5 M_\odot$, and fills
its Roche lobe. For such a binary with $M_s\simeq 0.5 M_\odot$,
we have $\R9\simeq 1$, $\D10\simeq 0.7$  and
$\Delta Q\simeq 5\times 10^{50}\Q52$, where
 $\Q52=Q/10^{52}$ ergs/st.
This can easily be more than the binding energy of the secondary
($GM_s^2/2R_s\approx 3 \times 10^{49}$ ergs)
(e.g., \cite{Shapiro83}; \cite{Nomoto82}).
Therefore, for strong GRBs the energy $\Delta Q$ suffices
 to completely evaporate the secondary. This is still true for a 
$1.4 M_\odot$ companion.
\par
The interaction between the relativistic GRB wind
 and the secondary is very complicated; it
depends, among other factors, on the properties 
of the winds. For the GRB model involving a strongly
magnetized, millisecond pulsar,
 the outflowing wind is dominated by a Poynting flux.
The luminosity in electron-positron pairs and
radiation is only $\sim 10^{-2}$ of the Poynting luminosity
 (e.g., \cite{Usov94}). The Lorentz factor of
the wind is $\sim 10^2-10^3$.
Typically, the diameter of the secondary white dwarf is rather smaller
than the thickness of the wind shell which is $\sim c\tau$, where
$\tau\sim 1-10$ s is the characteristic time of deceleration of the
pulsar rotation due to the action of the electromagnetic torque; this
is roughly the GRB duration or less.
The action of the relativistic, strongly magnetized
wind on the secondary
may be roughly modeled by assuming that the external pressure on its
 surface facing the GRB
increases instantly to

\begin{equation}
P_{\rm ext}\simeq {Q\over D^2c\tau}\simeq
3\times 10^{21}\Q52\D10^{-2}
\t1^{-1}\,\,\,{\rm ergs\,\,cm}^{-3}
\label{Pext}
\end{equation}
\noindent
for $\t1=\tau/1$ s $\sim 1-10$.
\par
Inside white dwarfs, electrons are free
and strongly degenerated (except for a very thin
surface layer with the density $\rho\lesssim 10^2$~g~cm$^{-3}$).
 At high density and low temperature,
these electrons give the main contribution
to the gas pressure irrespective of the element abundances.
The equation of state is (e.g., \cite{Shapiro83})

\begin{equation}
 P_e(\rho)\simeq 10^{23}\times\left\{
    \begin{array}{lcl}
     (\r6 /\me)^{5/3}\,\,\,
{\rm ergs\,\,cm}^{-3}\, ,
      &\ & {\rm for \,\,}
10^{-4}\lesssim \r6 < \me  ,\\
    (\r6/\me)^{4/3}\,\,\,
{\rm ergs\,\,cm}^{-3},
      &\ & {\rm for \,\,}
\r6 >\me,
    \end{array}
  \right.
  \label{P}
\end{equation}

\noindent
where $\r6=\rho/10^6\, {\rm g\, cm^{-3}}$, and
 $\me$ is the mean molecular weight per electron.
In white dwarfs helium, carbon, and oxygen dominate, so
$\me$ is nearly 2. The upper limit on the density for the validity of
equation (\ref{P}) is determined by neutronization,
 and, for example, for  helium it is $\r6\sim 10^5$.
\par
The instantaneous increase of external pressure
 from zero to $P_{\rm ext}$
results in formation of a strong
shock that propagates into the high density region.
Equations (\ref{Pext}) and (\ref{P}) imply that when 
the density in front of the shock is

\begin{equation}
\rho\lesssim \tilde\rho\simeq  10^5\Q52^{3/5}\D10^{-6/5}
\t1^{-3/5}\,\,\,{\rm g\,\,cm}^{-3}\,,
\label{tilderho}
\end{equation}
the pressure may be neglected.
In this case, the temperature behind the shock is about

\begin{equation}
\T9=T/10^9\,\, {\rm K}\simeq \Q52^{1/4}\D10^{-1/2}\t1^{-1/4}.
\label{T}
\end{equation}
For helium white dwarfs, this temperature may be higher than the
ignition temperature which varies from about $6\times 10^8$
K at $\rho \sim 10^5$~g~cm$^{-3}$ to $10^8$ K at
$\rho \sim 10^9$~g~cm$^{-3}$ (e.g., \cite{Nomoto82}). In this case,
the nuclear energy of the shocked matter is released
within a dynamical time scale, i.e., almost instantaneously.
At $\rho > \tilde\rho \sim 10^5$~g~cm$^{-3}$, the process
of thermonuclear burning propagates in the white dwarf
either as a supersonic detonation wave
or as a subsonic deflagration wave (\cite{Khokhlov93}
and references therein). It is worth noting that a transition from
a deflagration to a detonation is possible in the process of
burning propagation (e.g., \cite{Khokhlov97}).

For a helium secondary of reasonable mass
 (not too close to the Chandrasekhar
limit), the nuclear energy is enough for its complete disruption.
The detonation of such explosions is then similar in some respects 
to that of Type I supernovae (see below).
\par
The temperature given by equation
(\ref{T}) is at least a few times smaller than the ignition
temperature for carbon-oxygen mixtures
at $\rho \sim 10^5-10^6$~g~cm$^{-3}$
(e.g., \cite{Nomoto82}). However, an explosion might still occur
 for a carbon-oxygen WD if there is enough amplification of the 
inward shock due to the converging geometry of the phenomenon.
For an exactly spherical geometry the amplification is very large
(e.g., \cite{Zeldovich69}), but ours is only a semi-spherical 
implosion.

The equation of state for matter of hot white dwarfs is
(e.g., \cite{Cox68}; \cite{Shapiro83})

\begin{equation}
P_e(\rho, T)=P_e(\rho)+ \Delta P_e(\rho, T)\,,
\label{PrhoT}
\end{equation}
where $P_e(\rho)$ is the pressure of completely degenerated
electrons, given by equation (\ref{P}), and $\Delta P_e(\rho, T)$
is the thermal part of the electron
pressure $P_e(\rho, T)$. For $\rho > 10^6\me$~g~cm$^{-3}$,
we have

\begin{equation}
\Delta P_e(\rho, T)=2\left({\pi\over 3}\right)^{2/3}
\left({kT\over c\hbar}\right)^2
\left({m_p\me\over\rho}\right)^{2/3}P_e(\rho)\,,
\label{PT}
\end{equation}

\noindent
where $k$ is the Boltzmann constant, $c$ is the speed of light,
$\hbar$ is the Planck constant, and $m_p$ is the proton mass.

For $\me=2$, equations (\ref{P}) and (\ref{PT}) yield

\begin{equation}
\Delta P_e(\rho, T)\simeq
4\times 10^{22}\r6^{2/3}\T9^2\,\,
{\rm ergs\,\,cm}^{-3}\,.
\label{PTnum}
\end{equation}

Without detailed calculations  similar, e.g., to the two-dimensional
hydrodynamic simulations of supernova models  by \cite{Livne95} and
\cite{Livne99} we cannot tell whether, when the
disturbance reaches the center of the white dwarf, the temperature 
there is raised high enough for burning to occur there. 
We do not know how effective the 
shock convergence may be in amplifying the shock. 
We just parameterize the convergence effect by  
$\alpha$, the ratio of the thermal, electron
pressure reached at the center, $\Delta P_e(\rho, T)$, 
to that produced by the impact on the surface, $P_{\rm ext}$.
 Using  equations
(\ref{Pext}) and (\ref{PTnum}) we then have

\begin{equation}
\Tc9\simeq \left({\alpha\over 10}\right)^{1/2}
\Q52^{1/2}\D10^{-1}\t1^{-1/2}\rho_{c,6}^{-1/3},
\label{Tc}
\end{equation}

\noindent
where $\rho_c$ is the density at the stellar center.
For $\Q52\simeq 3$, $\D10\simeq 0.7$, $\t1\simeq 1$ and $\rc6\simeq 10$,
 from equation (\ref{Tc}) we can see that
for $\alpha > 20$ the temperature $T_c$ is higher than
the ignition temperature which is $\sim 1.7\times 10^9$~K
for carbon-oxygen mixtures at the density of $10^7$~g~cm$^{-3}$
(e.g., \cite{Nomoto82}). In this case explosion is expected to 
occur at the center.
\par
 Since detonation waves in WD matter have a finite
width it is important to compare this with the size of the star.
In carbon-oxygen WD matter the detonation wave
has roughly three spatially separated zones.
 The foremost is a sharp shock. The shock compresses and heats
 the material behind it, and carbon burning can start. This reaches 
a peak energy output within some distance $\Delta l_{_{\rm C}}$,
which typifies the carbon-burning zone. This scale
 roughly equals to the speed of
propagation of the shock multiplied by the carbon burning
time. The third zone is the region where
 matter is incinerated into nuclear-statistical-equilibrium
 (NSE) composition. The energy release in this layer is rather 
small, so it is not so important in the dynamics of the detonation
wave, but it is important in determining the composition of the 
ashes and the subsequent appearance of the explosion remnant.
\par
 For carbon-oxygen mixtures, the 
width of NSE-relaxation layer is many orders larger than the 
width of carbon burning, $\Delta l_{_{\rm NSE}}\gg 
\Delta l_{_{\rm C}}$. For $\rho_6\simeq 10$, we have
$\Delta l_{_{\rm C}}\simeq 1$~cm and $\Delta l_{_{\rm NSE}}
\simeq 10^8$~cm (e.g., \cite{Khokhlov89}). Since
$\Delta l_{_{\rm C}}\ll R_s$, the detonation wave may
form in the vicinity of the stellar center and
propagates to the surface. In the process of propagation 
of the outward detonation wave,  the width of carbon burning
is small, $\Delta l_{_{\rm C}}\ll R_s$,
except of a rather thin surface layer of the white dwarf, and
therefore most of the nuclear energy 
is released by the time the detonation wave reaches the surface.
 As noted above, the nuclear energy released 
is enough for the white dwarf to be completely disrupted.
\par
The kinetics of helium burning differs substantially from that 
of carbon-oxygen burning. The leading reaction is
 $3\, ^{4}$He $\rightarrow ^{12}$ C.
At the same densities, the rate of this reaction is much smaller 
than that of carbon burning. As a result the  width of 
the nuclear-burning zone is many orders larger (e.g., \cite{Khokhlov89}). 
Still, for $\rho_6\sim 0.1-1$ the width of the helium-burning zone
 is small enough, $\Delta l_{_{\rm He}}
\simeq10^8$~cm $\ll R_s$, so that we can take this zone as small
 compared with the other relevant scales. Thus, an inward 
detonation wave can lead to the white dwarf explosion 
as we discussed above.
\par
For rather massive white dwarfs, $M_s\gtrsim 0.8M_\odot$, which 
can undergo accretion-induced explosions, the mean density of 
the bulk matter is $\gtrsim 10^6$~g~cm$^{-3}$, and 
the NSE relaxation width is small, $\Delta l_{_{\rm NSE}}\ll R_s$.
In this case, incineration is 
effective, and the remaining ashes consist mostly of $^{56}$Ni,
which decays and provides energy for the long-time radiation
of Type I supernovae (e.g. \cite{Nomoto82}; \cite{Woosley86}).
In contrast, for low-mass white dwarfs, $M_s\lesssim 0.3M_\odot$,
which cannot be exploded by gas accretion without 
significant increase of their masses, the great fraction of the mass
is at lower densities ($\rho \lesssim 10^5$~g~cm$^{-3}$) for 
which $\Delta l_{_{\rm NSE}}\gg R_s$ and the production of 
$^{56}$Ni is strongly suppressed irrespective the
abundance (\cite{Khokhlov89}; \cite{Nomoto82}; \cite{Woosley86}).
Therefore, for such a low-mass white dwarf
the mass of $^{56}$Ni that is produced in the 
GRB-induced explosion is very low, and
this explosion leads to a weak supernova-like
phenomenon, which differs qualitatively from known
supernovae. Observation of such a phenomenon correlated with GRBs
could confirm our idea on the GRB-induced explosions of secondary
white dwarfs. 

\section{Discussion}

It is generally believed that Type I supernovae are produced by
thermonuclear explosions of white dwarfs (e.g.,
\cite{Nomoto82}; \cite{Woosley86}; \cite{Niemeyer97}).
Such explosions may be brought about by accretion of matter
onto the white dwarfs. In the process,
thermonuclear burning that is triggered near the surface propagates
either in the form of a supersonic detonation wave,
or as a subsonic deflagration. This
results in the incineration of most of the white dwarf matter into
$^{56}$Ni, which is ejected from the star. The radioactive decay
$^{56}$Ni $\rightarrow$ $^{56}$Co $\rightarrow$ $^{56}$Fe
can provide a sufficient amount of late-time energy input to
power the light curves of Type I supernovae.
\par
In this paper we have argued that a similar fate 
may befall a white dwarf that is exposed 
to the ejecta of a GRB explosion in a very close binary 
companion. The GRB angular energy output, $Q$, near the binary plane
which is necessary for the WD explosion is $\sim 10^{52}$ ergs/st.
This is a typical value of $Q$ in the model of GRBs we considered.    
Indeed, the rotational energy of millisecond pulsars that
is a plausible source of energy for GRBs may be 
as high as a few $\times 10^{53}$ ergs, and almost all this energy
may be transformed into the energy of a relativistic, strongly 
magnetized wind (e.g., \cite{Usov94}). 
The angular distribution of the wind flux depends on 
the angle $\vartheta$ between the rotational and magnetic axes 
and varies within a factor of 2-3 or so.
Such a moderate collimation of the outflowing wind may be 
either along the rotational axes of the pulsar at $\vartheta \simeq 0$
(\cite{Benford84}; \cite{Michel85}) or near the equator at
$\vartheta\simeq \pi/2$ (\cite{Belinsky94}). 
For a very close WD + WD binary that is the predecessor of 
the GRB source, one expects the secondary white dwarf to be 
near the equator of the millisecond pulsar which forms by 
accretion-induced collapse of the primary white dwarf. 
In this case, the $Q$ value in the WD direction 
is typically $\sim 10^{52}-10^{53}$ ergs/st, 
the maximum value of it being reached when both the pulsar rotation
is extremely fast and the magnetic axis of the pulsar
is perpendicular to its rotational axis.

We suggest that the resulting explosion is similar in some respects to Type I
supernova, but may differ substantially in others,
 especially if the mass of the
WD is small, $M_s\lesssim 0.3M_\odot$.
First, because the trigger mechanism is different, a different elemental
 abundance may result. Second, the post-explosion WD remnant 
has ample time ($\sim 10$ s or more) to interact  with the
relativistic wind outflowing from the GRB source.
This can lead to additional acceleration of the
explosion debris even, for some parts, up to relativistic velocities. 
Therefore, for GRB-induced supernovae the maximum 
of their light curves is expected to be observed substantially earlier 
than for typical Type I supernovae. Taking also into account that typically
the amount of radioactive $^{56}$Ni produced in GRB-induced supernovae
is low, the luminosities of these supernovae may
decrease fast after the maximum without long lived tails.

At present, several SN/GRB associations have been suggested (for a review,
see \cite{Wheeler99}). Among them, SN 1998bw, possibly associated with 
GRB 980425 (\cite{Galama98}), is the most famous and best established 
candidate for such an association. 
SN 1998bw that was very powerful could not have been produced
by our mechanism, which produces rather weak optical supernovae.
The amount of radioactive $^{56}$Ni produced in
SN 1998bw has been estimated to be $\sim 0.5-0.75M_\odot$ 
(e.g., \cite{Iwamoto98}; \cite{Woosley99}), much more than the explosion
we discuss can make. Also, 
if GRB 980425 is connected with SN 1998bw its total energy released,
even if it is isotropic, is only $\sim 10^{48}$ ergs. This is about four 
orders less than what is necessary for induction of a WD explosion.
And third, the optical properties of SN 1998bw indicate that the 
progenitor star (like the progenitor stars of all other supernovae possibly
associated with GRBs) was a massive star with a mass at
least a several times larger that the maximum possible mass of WDs
(\cite{Iwamoto98}; \cite{Woosley99}). While the observations of
GRB 980425-SN 1998bw may be explained fairly well in the collapsar
model (e.g., \cite{MW99}), we suggest that
at least some cosmological GRBs may be associated with rather weak
supernova-like explosions of low-mass $(\lesssim 0.3-0.5M_\odot)$
white dwarfs. In our scenario, GRBs and SNs associated with each other 
are different phenomena while in the collapsar model the two events are one. 

Recently, possible evidence for the existence of iron K-shell emission
lines has been found in two GRBs:
GRB 970508 and GRB 970828 (\cite{Piro99}; \cite{Yoshida99}).
The presence of dense matter very close ($\lesssim 10^{16}$ cm) to the 
GRBs, which is not expanding relativistically, is required by these
 observations. 
The remnants of the GRB-induced explosions of white dwarfs may 
be responsible for emission of the iron lines. For this, it is necessary
that a small fraction of the remnant matter with iron mass of $\sim 10^{-5}-
10^{-4}M_\odot$ is accelerated by the GRB wind to subrelativistic velocities 
 and generates the Fe line emission at the distance of $\sim (1-3)\times 
10^{15}$ cm from the GRB source. It is worth noting that
rather strong, high-redshift ($z\gtrsim 1$)
GRBs, like GRB 970828, with X-ray afterglows (for their prompt localization)
and without standard optical afterglow  are the best candidates
for searching the possible, weak, supernova-like explosions posited here.
\par
Induction of supernova-like explosions by GRBs
is similar in many respects to  ablation in laser and heavy-ion fusion
(for review, see \cite{Meyer98}). It is well-known that the symmetry of 
the irradiation of the fusion fuel is crucial for its successful explosion.
Otherwise, only the outer layers may be affected. The same may be true for
carbon-oxygen WDs if the driving external pressure $P_{\rm ext}$ is not
spherical enough. Other obstacles to successful
explosion might result from  numerous instabilities that may 
develop at the  surface. In our case, though, plasma instabilities 
at the  surface  where the GRB wind interacts with the WD matter
may be suppressed by a very strong magnetic field of the GRB wind.

\acknowledgments

We thank an anonymous referee for useful suggestions.
This research was supported by the MINERVA Foundation, Munich, Germany.


\newpage

\end{document}